# Decentralised Trust for the Digital Economy

Geoffrey Goodell, University College London (g.goodell@ucl.ac.uk)

We propose a research initiative to explore and evaluate end-user technology, infrastructure, business imperatives, and regulatory policy to support the privacy, dignity, and market power of individual persons in the context of the emerging digital economy. Our work shall take a "system-level" approach to the design of technology and policy, considering the outcomes associated with the implementation and deployment of systems consisting of operational infrastructure, policies, and protocols for humans and computers alike. We seek to define and evaluate a set of approaches to the design and implementation of systems whose features specifically support the rights and market power of individual persons and local organisations, for the explicit goal of supporting truly consensual trust relationships and empowering local communities and organisations.

## SUMMARY

Individuals who use communication platforms for messaging and social media, as well as infrastructure services such as transportation and government services, leave a data trail that in many cases results in a permanent record of their activities. This is problematic for several reasons. It exposes them to manipulation, control, and exploitation through surveillance. It creates a permanent record of their habits and activities that they can never escape, creating a chilling effect for legitimate but unconventional behaviours. It provides a means by which providers of services can discriminate against them on the basis of their established profiles.

We consider how the digital economy might evolve in the presence of technology for the collection, aggregation, and analysis of data concerning personal transactions and attributes. The functionality offered by new data-intensive technology demonstrates that platform-based services that compromise the interests of individual persons are often more efficient than their alternatives, in turn demonstrating a public need to explicitly support systems that offer privacy and fairness by design.

Our research programme seeks to develop purpose-built technology architectures that enable individual persons and retail consumers to communicate privately and develop trust relationships on their own terms. We shall focus not only on the technical and operational feasibility of such systems but the broader regulatory, business, and policy context as well, with close attention to the interface between the technical systems we envision and their interfaces to governments and institutions, including but not limited to industry participants and marketplaces.

## BACKGROUND AND MOTIVATION

Human autonomy is increasingly at risk. Systems and devices that people use every day collect, aggregate, and analyse data about their habits, financial circumstances, travel, and relationships, creating a detailed and permanent record of their lives. Such records are exchanged by data brokers and exploited to control populations cheaply and at scale. The engineering research community can address this problem with privacy-enabling technologies, not only at the mathematical level with cryptography and the technical level with privacy tools and anonymity systems, but at the broader system level with institutional architectures that achieve the goals of governments and businesses without requiring the collection of data that can be used to undermine the interests of individual persons. Research is greatly needed to achieve security, efficiency, and economic suitability of such systems and to address the human-computer interaction requirements that make such systems appropriate and usable by the wider population.

Many activities requiring interaction with infrastructure are highly sensitive to the autonomy and privacy of individuals and businesses. New technology has emerged that promotes centralised surveillance and influence over these activities through default linkage of individuals' identities by state and non-state actors. The working group shall consider how legitimate institutional and regulatory interests can be achieved without compromising the legitimate interest of individuals or businesses in protecting their autonomy and privacy. The group shall also consider how trust can be both protected and enhanced in the consent mechanisms for access to identity information and advance proposed solutions. It is observed that:

- systematic collection of data of individual persons by corporations and state actors introduces a means of influence that undermines individuals' interests as well as their ability to act autonomously, and therefore impacts trust in democratic institutions and firms;

- regulation of infrastructure including but not limited to payment channels, border control, communication, transportation, and marketplaces generally requires extensive collection of data, but does not require systematic surveillance of individuals to achieve its primary objectives; and

- technology and policy must be established via a coordinated effort, acknowledging that it is insufficient to consider only existing technology when establishing new policy or to consider only existing policy when building new technology.

Borrowing from the discussion in our earlier work "A Decentralised Digital Identity Architecture" (see **Related Articles**), we specifically consider the increasingly commonplace application of identity systems "to facilitate targeting, profiling and surveillance" by "binding us to our recorded characteristics and behaviours" [1]. Although we focus primarily upon the application of systems for digital credentials to citizens of relatively wealthy societies, we hope that our proposed architecture might contribute to the identity zeitgeist in contexts such as humanitarian aid, disaster relief, refugee migration, and the special interests of children as well.

We argue that while requiring strong non-transferability might be appropriate for some applications, it is inappropriate and dangerous in others. Specifically, we consider the threat posed by mass surveillance of ordinary persons based on their habits, attributes, and transactions in the world. Although the governments of Western democracies might be responsible for some forms of mass surveillance, for example via the recommendations of the Financial Action Task Force [2] or various efforts to monitor Internet activity [3, 4], we note that the siren song of surveillance capitalism [5], including the practice of "entity resolution" through aggregation and data analysis [6], presents a particular risk to human autonomy. We suggest that many "everyday" activities such as normal retail remittances, the use of library resources, public transportation services, and mobile data services are included in a category of activities for which strong non-transferability is not necessary and for which there is a genuine need for technology that explicitly protects the legitimate privacy interests of individual persons.

**RESEARCH QUESTIONS**

Our research topics for this effort shall include questions such as:

1. **What are the set of design requirements and protocols needed to provide a means by which individuals and groups can engage in peer-to-peer communication without exposing metadata about their identities or relationships to third parties?** Much work has been done on decentralised identifiers (DIDs) and anonymous communication, although bootstrapping such systems is difficult given the role of network carriers and device platforms in blocking truly peer-to-peer communication. How can we enable such communication, including pairwise communication, introductions, and group broadcast, within the current set of real-world constraints, without requiring ordinary users and small groups to operate their own servers and expensive platform infrastructure?

2. **What user design features are needed to allow individuals to establish multiple identities that they can use to establish separate and distinct relationships with counterparties, including other individuals, businesses, and governments?** Many popular paradigms such as 'single sign on' implicitly encourage individuals to aggregate their relationships into a single point of control. However, human relationships are more complex, mediated by the individuals themselves. What design features are needed to place individual persons at the centre of the relationships they form, in a way that they can choose to selectively bridge features of different relationships and retain control over what they reveal? What systems would be needed to allow businesses and governments to interoperate with such identities?

3. **How do technical limitations and user experience requirements impact the design of regulated systems that support accountless access by consumers, and what design tradeoffs are involved?** We have observed that blind signatures can allow individuals to anonymously make certified claims about themselves. We have shown how distributed ledgers can be used to support the claims made by a user while limiting the channels by which the user might be tracked. We seek to understand the various ways in which the architecture we proposed for anonymous, unlinked credentials might be implemented, including a taxonomy of use cases and specifically how the technical tradeoffs might apply to different use cases.

4. **What are the appropriate use cases for systems that use anonymous tokens without strong non-transferability, and what are the limits to the applicability of such systems?** Anonymous tokens without strong non-transferability might be needed to protect privacy rights in the context of everyday activities such as routine retail purchases, use of public transportation, and so on. Even if this is enough to warrant the deployment of such a system, what are the limits to its applicability? For example, would it be necessary to limit the volume of withdrawals per unit of time from a regulated account into a cash-like store of value? Would anonymous credentials not work for some use cases, such as crossing international borders? Where would the lines be drawn, and would a layered approach offer a way to achieve the necessary security without undermining privacy in the routine cases?

5. **What policy and governance mechanisms would be needed to support systems that use anonymous tokens representing value or credentials, without assuming universally trusted operators or service providers?** What would be the protocols by which the various providers of services enforce the rules of the system, interact with users, and invite additional participants? Would it be possible to expect private industry to operate such a system of its own accord, or would government participation would be needed? If government involvement is necessary, then would a co-regulation model apply? How would we ensure that the rights of individual persons are protected?

6. **What are the design considerations for a marketplace for retail goods and services that does not rely upon a long-lived relationship between an individual and a market platform?** Would it be possible to develop a co-regulatory framework for dealers of retail products and services of the sort that are often used for dealers of securities, and if so, then who would be the participants? What are the requirements for fungibility and liquidity for products and services traded in a best execution network, and how could markets for retail products and services be adapted to meet these requirements?

**OBJECTIVES**

In addition to answering the specific research questions enumerated above, we seek to steer the academic and public dialogue toward a systems-level approach that considers technology and policy as component elements of a solution that intends to deliver particular outcomes. We specifically consider the following requirements for the design of infrastructure systems:

1. Minimise **control points** that can be used to co-opt the system.

2. Mitigate architectural characteristics that lead to **mass surveillance** of individual persons.

3. Do not impose **non-consensual** trust relationships upon beneficiaries.

4. Empower individual users to manage the **linkages** among their activities.

We specifically consider systems of the future with respect to the three streams: remittances, access to infrastructure services, and digital marketplaces. We seek to elaborate legal, business, and technical requirements for the design of infrastructure systems toward identifying which requirements, if any, truly are truly incompatible with the right to private use of the system. In addition, we seek to elaborate the manner by which the requirements that are not incompatible with privacy rights influence the design tradeoffs for the system, and we seek to offer privacy-preserving technical designs and compatible policy frameworks that could potentially work within the existing legal and business environment.

**APPROACH**

We intend to conduct a set of research projects that explore and evaluate privacy-enabling polices, conventions, and technologies to support a variety of personal and commercial interactions with the digital economy, including but not limited to electronic payments, financial clearing and settlement, and the physical movement of people, goods, and services. We envision our role as part of a broader community effort, in which we specifically shall undertake the following:

1. Conduct academic research on the interface of policy and technology in the domain of anonymity and the digital economy, and the consent mechanisms and good practices needed to protect the privacy interests of individuals and businesses;

2. Support the development of prototypes and 'proofs-of-concept' to solve real-world problems faced by institutions and consumers;

3. Participate in the development of national and international standards related to business practices and technology architecture for the purpose of promoting interoperability; and

4. Engage with government and industry groups toward securing the institutional adoption of privacy-enabling infrastructure and business practices, with particular consideration of identity, consent, and verifiability of claims.

We thank Peer Social Foundation for its sponsorship of this work. We anticipate that this project and its objectives will evolve over time in accordance with the mutual interests of its participants. We thank Paul Ferris and Professor Tomaso Aste for their contributions to earlier versions of this document. Geoff Goodell is also an associate of the Systemic Risk Centre of the London School of Economics as well as the UCL Centre for Blockchain Technologies. We also thank the European Commission for the FinTech project (H2020-ICT-2018-2 825215).

**RELATED ARTICLES**